\documentclass[aps,prb,twocolumn,amssymb,showpacs,amsmath,superscriptaddress,floatfix]{revtex4-1}
\usepackage{graphicx}
\usepackage{bm}
\begin{document}

\title {Interplay between Chiral Charge Density Wave and Superconductivity in Kagome Superconductors: A Self-consistent Theoretical Analysis}

\author{Hong-Min Jiang}
\email{monsoonjhm@sina.com} \affiliation{School of Science, Zhejiang
University of Science and Technology, Hangzhou 310023, China}
\author{Min Mao}
\affiliation{School of Science, Zhejiang University of Science and
Technology, Hangzhou 310023, China}
\author{Zhi-Yong Miao}
\affiliation{School of Science, Zhejiang University of Science and
Technology, Hangzhou 310023, China}
\author{Shun-Li Yu}
\email{slyu@nju.edu.cn} \affiliation{National Laboratory of Solid
State Microstructures and Department of Physics, Nanjing University,
Nanjing 210093, China} \affiliation{Collaborative Innovation Center
of Advanced Microstructures, Nanjing University, Nanjing 210093,
China}
\author{Jian-Xin Li}
\email{jxli@nju.edu.cn} \affiliation{National Laboratory of Solid
State Microstructures and Department of Physics, Nanjing University,
Nanjing 210093, China} \affiliation{Collaborative Innovation Center
of Advanced Microstructures, Nanjing University, Nanjing 210093,
China}

\date{\today}

\begin{abstract}
Inspired by the recent discovery of a successive evolutions of
electronically ordered states, we present a self-consistent
theoretical analysis that treats the interactions responsible for
the chiral charge order and superconductivity on an equal footing.
It is revealed that the self-consistent theory captures the
essential features of the successive temperature evolutions of the
electronic states from the high-temperature ``triple-$Q$" $2\times
2$ charge-density-wave state to the nematic charge-density-wave
phase, and finally to the low-temperature superconducting state
coexisting with the nematic charge density wave. We provide a
comprehensive explanation for the temperature evolutions of the
charge ordered states and discuss the consequences of the
intertwining of the superconductivity with the nematic charge
density wave. Our findings not only account for the successive
temperature evolutions of the ordered electronic states discovered
in experiments but also provide a natural explanation for the
two-fold rotational symmetry observed in both the
charge-density-wave and superconducting states. Moreover, the
intertwining of the superconductivity with the nematic charge
density wave order may also be an advisable candidate to reconcile
the divergent or seemingly contradictory experimental outcomes
regarding the superconducting properties.
\end{abstract}

\pacs{74.20.Mn, 74.25.Ha, 74.62.En, 74.25.nj}
\maketitle

\section{introduction}

Kagome systems, with their geometrical frustration and nontrivial
band topology, have long served as paradigmatic platforms for
investigating exotic quantum phases of electronic matter, including
spin liquid~\cite{YRan11,HCJiang11,SYan1,
StefanDepenbrock11,TianHengHan11,YinChenHe11,HJLiao11,ZiliFeng11,PKhuntia11},
various topological quantum
phases~\cite{HMGuo1,SLYu11,JunWen11,ETang11,GXu11,LindaYe11,DFLiu11,JiaXinYin11},
charge density wave (CDW)~\cite{HMGuo1,GAFiete11}, spin density
wave~\cite{SLYu1}, bond density wave~\cite{Isakov1,Kies2,WSWang1}
and superconductivity~\cite{SLYu1,Kies2,WSWang1,WHKo1,JKang11}. Of
particular interest is the possible phases near the van Hove filling
(VHF), especially the superconducting (SC) state, where the
density-of-states (DOS) is extremely enhanced and the Fermi surface
(FS) exhibits perfect nesting~\cite{SLYu1}. These unique properties
of the electron structure lead to the SC state being susceptible to
competition from various other electronic
instabilities~\cite{SLYu1,WSWang1,Kies2}. Understanding the
superconductivity in such a kagome material that either avoids or
even intertwines with these competing instabilities remains an
unsettled issue.

The recent discovery of superconductivity in a family of compounds
AV$_{3}$Sb$_{5}$ (A=K, Rb, Cs), which share a common lattice
structure with kagome net of vanadium atoms, has set off a new boom
of researches on the
superconductivity~\cite{Ortiz1,SYYang1,Ortiz2,QYin1,KYChen1,YWang1,ZZhang1,YXJiang1,
FHYu1,XChen1,HZhao1,HChen1,HSXu1,Liang1,CMu1,CCZhao1,SNi1,WDuan1,
Xiang1,PhysRevX.11.041030,PhysRevB.104.L041101,
PhysRevX.11.041010,NatPhys.M.Kang,YFu1,HTan1,Shumiya1,FHYu2,LYin1,Nakayama2,
KJiang1,LNie1,HLuo1,Neupert1,YSong1,Nakayama1,HLi1,CGuo1,HLi2,HLi3,XWu1,Denner1,SCho1,YPLin3,
LZheng1,CWen1,Tazai1,Jiang1,ZLiu1,Mielke1}. The appealing aspects of
these compounds lie in that they incorporate many remarkable
properties of the electron structure, such as VHF, FS nesting and
nontrivial band topology~\cite{Ortiz1}. Consistent with the fairly
good FS nesting and proximity to the von Hove singularities, the
system undergoes a ``triple-$Q$" $2\times 2$ CDW transition at
temperature $T_{CDW}\sim 78-104K$, with the in-plane wave vectors
align with those connecting the van Hove
singularities~\cite{Ortiz1,YXJiang1,HZhao1,Liang1,HChen1,HLi2,Mielke1}.
While the neutron scattering~\cite{Ortiz3} and muon spin
spectroscopy~\cite{Kenney1} measurements have ruled out the
possibility of long-range magnetic order in AV$_{3}$Sb$_{5}$, a
significant anomalous Hall effect is still observed above the onset
of the SC state in this $2\times 2$ CDW phase~\cite{SYYang1,FHYu1},
indicating a time-reversal symmetry-breaking state originating from
the charge degree of freedom~\cite{Kenney1}. So far, there are an
increasing number of experimental evidences supporting that the CDW
state has a $2\times 2$ chiral flux
order~\cite{YXJiang1,Shumiya1,CGuo1,Mielke1,LYu1,YXu1,XZhou1,DChen1,YHu1},
i.e., the chiral flux phase (CFP)~\cite{Feng1,Feng2,JWDong1}.
Furthermore, the muon spin relaxation technic observed a noticeable
enhancement of the internal field width, which takes place just
below the charge ordering temperature and persists into the SC
state~\cite{Mielke1}, suggesting an intertwining of time-reversal
symmetry breaking charge order with superconductivity.

Nevertheless, more recent experiments revealed that the
high-temperature $2\times 2$ CDW state does not directly border the
low temperature SC state. Instead, the high-temperature $2\times 2$
CDW state is separated from the SC ground state by an
intermediate-temperature regime with the two-fold ($C_{2}$)
rotational symmetry of electron
state~\cite{HLi1,ZJiang1,LNie1,PWu1}. This electronic state with
$C_{2}$ rotational symmetry is found to appear at temperature
$T_{nem}$ well below $T_{CDW}$ and persist into the SC state, as
evidenced by transport~\cite{Xiang1} and scanning tunneling
microscopy (STM) measurements~\cite{HZhao1,HLi1}.

Apart from the exotic charge orders, the superconductivity in
AV$_{3}$Sb$_{5}$ exhibits some unusual features as well. On the one
hand, the SC pairings in these compounds are suggested to be of the
$s$-wave type, supported by the appearance of the Hebel-Slichter
coherence peak just below $T_{c}$ in the nuclear magnetic resonance
spectroscopy~\cite{CMu1} and the nodeless SC gap in both the
penetration depth measurements~\cite{WDuan1} and the angle-resolved
photoemission spectroscopy (ARPES) experiment~\cite{YZhong1}. On the
other hand, the indications of time-reversal symmetry breaking and
the $C_{2}$ rotational symmetry discovered in the SC
state~\cite{Mielke1,Xiang1,HZhao1,HLi1,YZhong1}, together with the
nodal SC gap feature detected by some
experiments~\cite{CCZhao1,HSXu1,Liang1,HChen1}, hint to an
unconventional superconductivity.

Since the superconductivity occurs within the density wave ordered
state, understanding the relationship between the CDW instability
and superconductivity is a central issue in the study of
AV$_{3}$Sb$_{5}$. Theoretical analysis has shown that a conventional
fully gapped superconductivity is unable to open a gap on the
domains of the CFP and results in the gapless edge modes in the SC
state~\cite{YGu1}. A more direct consideration of the impact of the
chiral $2\times 2$ CDW on the SC properties has revealed that a
nodal SC gap feature shows up even if an on-site $s$-wave SC order
parameter is included in the study~\cite{HMJiang2}. However, there
is still limited knowledge about the rotational symmetry breaking
phase that straddles the SC ground state and the $2\times 2$ CDW
state in this class of kagome metals, particularly its origin, its
role in the formation of the superconductivity, and its impact on
the SC properties.

In this paper, we investigate the interplay between the CFP and
superconductivity in a fully self-consistent theory, which
self-consistently treats both the chiral CDW and the SC pairing
orders on an equal footing. The calculated results catch the
essential characteristics of the successive temperature evolutions
of the electronically ordered states, starting from the
high-temperature $2\times 2$ ``triple-$Q$" CFP (TCFP) to the nematic
CFP (NCFP), and finally to the low-temperature SC state. Notably,
the SC state emerges in the coexistence with the NCFP, by which the
free energy in the coexisting phase is significantly lowered than
that in the pure SC state. The rotational symmetry-breaking
transition of the CDW can be understood from a competitive scenario,
in which the delicate competition between the doping deviation from
the VHF and the thermal broadening of the FS determines the
energetically favored state. In the coexisting phase of the $s$-wave
SC pairing and the NCFP order, the DOS exhibits a nodal gap feature
manifesting as the V-shaped DOS along with the residual DOS near the
Fermi energy. These results not only reproduce the successive
temperature evolutions of the ordered electronic states observed in
experiment, but also provide a tentative explanation to the two-fold
rotational symmetry observed in both the CDW and SC states.
Furthermore, the intertwining of the SC pairing with the NCFP order
may also be an advisable candidate to reconcile the divergent or
seemingly contradictory experimental outcomes concerning the SC
properties.

The remainder of the paper is organized as follows. In Sec. II, we
introduce the model Hamiltonian and carry out analytical
calculations. In Sec. III, we present numerical calculations and
discuss the results. In Sec. IV, we make a conclusion.

\section{model and method}
\begin{figure}
\begin{center}
\vspace{.2cm}
\includegraphics[width=230pt,height=200pt]{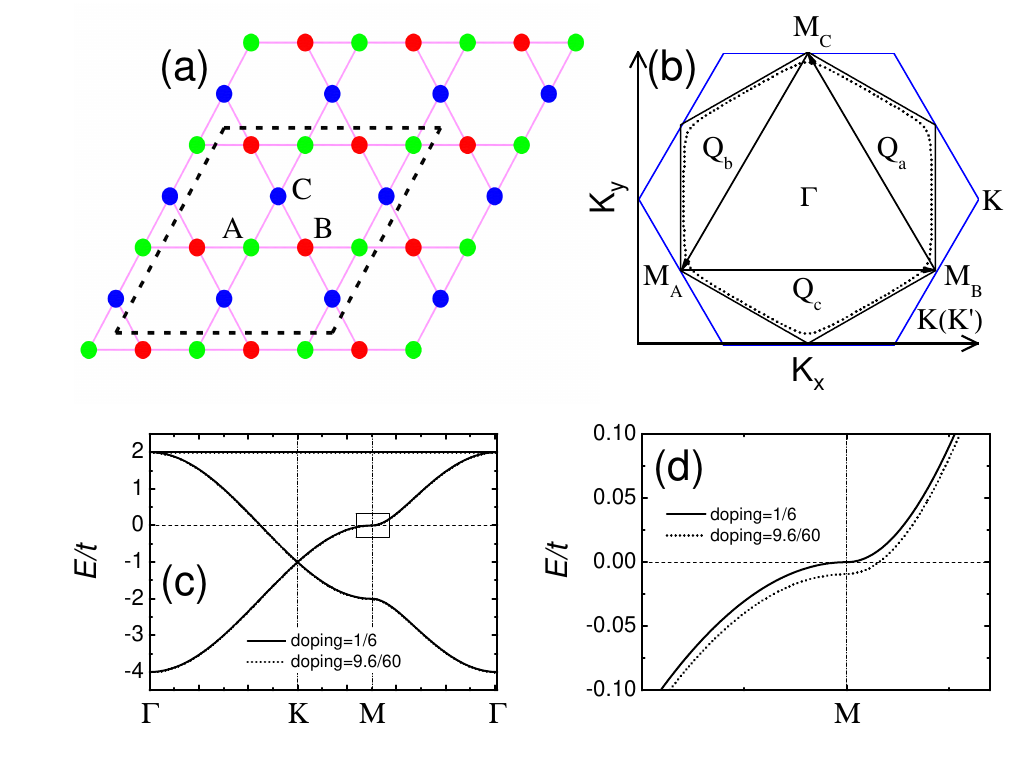}
\caption{(a) Structure of the kagome lattice, made out of three
sublattices $A$ (green dots), $B$ (red dots) and $C$ (blue dots).
The dashed lines in the figure denote the enlarged unit cell in the
$2\times 2$ charge density wave state. (b) Fermi surface produced by
the Hamiltonian $H_{0}(\mathbf{k})$ for the doping levels $1/6$
(solid lines) and $9.6/60$ (dotted lines), respectively. (c)
Tight-binding dispersion along high-symmetry cuts for doping levels
1/6 (solid curve) and 9.6/60 (dotted curve) (Note that the solid and
dotted curves are close to each other, so only one solid curve can
be discerned in a large energy scale). The dashed line denotes the
Fermi level. (d) The enlarged view of the square box in (c).}
\label{fig1}
\end{center}
\end{figure}

It is generally considered that the scattering due to the FS
nesting, especially the inter-scattering between three van Hove
points with the nesting wave vectors
$\mathbf{Q}_{a}=(-\pi,\sqrt{3}\pi)$,
$\mathbf{Q}_{b}=(-\pi,-\sqrt{3}\pi)$ and $\mathbf{Q}_{c}=(2\pi,0)$
shown in Fig.~\ref{fig1}(b), is closely related to the CDW in
AV$_{3}$Sb$_{5}$. Meanwhile, the VHF was also proposed to be crucial
to the superconductivity in AV$_{3}$Sb$_{5}$. A single orbital tight
binding model near the VHF produces the essential feature of the FS
and the van Hove physics~\cite{SLYu1}. Therefore, to capture the
main physics of the chiral CDW and its intertwining with the SC in
AV$_{3}$Sb$_{5}$, we adopt a minimum single orbital model. We also
note that the six-fold ($C_{6}$) symmetry is broken within the
unit-cell of the $2\times 2$ CDW state~\cite{LNie1}, without any
additional reduction in translation symmetry. Thus, we choose the
enlarged unit cell (EUC) with size $2\mathbf{a}_{1}\times
2\mathbf{a}_{2}$, as indicated by the dashed lines in
Fig.~\ref{fig1}(a).

The single orbital model can be described by the following
tight-binding Hamiltonian,
\begin{eqnarray}
H_{0}&=&-t\sum_{\langle
\textbf{ij}\rangle\alpha}c^{\dag}_{\textbf{i}\alpha}c_{\textbf{j}\alpha}
-\mu\sum_{\textbf{i}\alpha}c^{\dag}_{\textbf{i}\alpha}c_{\textbf{i}\alpha},
\end{eqnarray}
where $c^{\dag}_{\textbf{i}\alpha}$ creates an electron with spin
$\alpha$ on the site $\mathbf{r}_{i}$ of the kagome lattice and
$\langle\textbf{ij}\rangle$ denotes nearest-neighbors (NN). $t$ is
the hopping integral between the NN sites, and $\mu$ stands for the
chemical potential. The Hamiltonian $H_{0}$ can be written in the
momentum space as,
\begin{eqnarray}
H_{0}(\mathbf{k})&=&\sum_{\mathbf{k}\alpha}\hat{\Psi}^{\dag}_{\mathbf{k}\alpha}\hat{\mathcal{H}}^{0}_{\mathbf{k}}
\hat{\Psi}_{\mathbf{k}\alpha},
\end{eqnarray}
with
$\hat{\Psi}_{\mathbf{k}\alpha}=(c_{A\mathbf{k}\alpha},c_{B\mathbf{k}\alpha},c_{C\mathbf{k}\alpha})^{T}$
and
\begin{eqnarray}
\hat{\mathcal{H}}^{0}_{\mathbf{k}}=\left(
\begin{array}{ccc}
-\mu & -2t\cos k_{1} &
-2t\cos k_{2} \\
-2t\cos k_{1} & -\mu & -2t\cos k_{3} \\
-2t\cos k_{2} & -2t\cos k_{3} & -\mu
\end{array}
\right).
\end{eqnarray}
The index $m=A,B,C$ in $c_{mk\sigma}$ labels the three basis sites
in the triangular primitive unit cell (PUC) as shown in
Fig.~\ref{fig1}(a). $k_{n}$ is abbreviated from
$\mathbf{k}\cdot\mathbf{\tau}_{n}$ with
$\mathbf{\tau}_{1}=\hat{x}/2$,
$\mathbf{\tau}_{2}=(\hat{x}+\sqrt{3}\hat{y})/4$ and
$\mathbf{\tau}_{3}=\mathbf{\tau}_{2}-\mathbf{\tau}_{1}$ denoting the
three NN vectors. The spectral function of $H_{0}(\mathbf{k})$
defined as
$A^{0}(\mathbf{k},E)=-\frac{1}{\pi}\textmd{Tr}[\textmd{Im}\hat{G}^{0}(\mathbf{k},iE\rightarrow
E+i0^{+})]$ with
$\hat{G}^{0}(\mathbf{k},iE)=[iE\hat{I}-\hat{\mathcal{H}}^{0}_{\mathbf{k}}]^{-1}$.
Near the VHF with $1/6$ hole doping, the Hamiltonian
$\hat{\mathcal{H}}^{0}_{\mathbf{k}}$ generates the hexagonal FS and
the corresponding energy band as shown in Figs.~\ref{fig1}(b) and
(c) respectively, which capture the essential features of the FS and
energy band observed in the ARPES experiment and the density
functional theory calculations~\cite{Ortiz1}.

The second part of the Hamiltonian incorporates the orbital current
order,
\begin{eqnarray}
H_{C}=&&\sum_{\langle\textbf{ij}\rangle
\alpha}iW_{\mathbf{ij}}c^{\dag}_{\mathbf{i}\alpha}c_{\mathbf{j}\alpha},
\end{eqnarray}
where
$W_{\mathbf{ij}}=-V_{c}\textmd{Im}\langle\chi_{\mathbf{ij}}^{\dag}\rangle$
denotes the mean-field value of the magnitude of the orbital current
order in the CFP with
$\chi_{\mathbf{ij}}=c^{\dag}_{\mathbf{i}\uparrow}c_{\mathbf{j}\uparrow}
+c^{\dag}_{\mathbf{i}\downarrow}c_{\mathbf{j}\downarrow}$. The
orbital current order could be derived from the Coulomb interaction
between electrons on neighboring sites, i.e.,
$H_{V}=V_{c}\sum_{\mathbf{ij}}n_{\mathbf{i}}n_{\mathbf{j}}$ with
$n_{\mathbf{i}}=\sum_{\alpha}c^{\dag}_{\mathbf{i}\alpha}c_{\mathbf{i}\alpha}$.
A straightforward algebra shows that
\begin{eqnarray}
n_{\mathbf{i}}n_{\mathbf{j}}=&&2n_{\mathbf{i}}-
\sum_{\alpha\beta}(c_{\mathbf{i}\alpha}^{\dag}c_{\mathbf{j}\beta})
(c_{\mathbf{i}\alpha}^{\dag}c_{\mathbf{j}\beta})^{\dag} \nonumber\\
=&&2n_{\mathbf{i}}-\frac{1}{2}
\sum_{\eta=0}^{3}S_{\mathbf{ij}}^{\eta}(S_{\mathbf{ij}}^{\eta})^{\dag},
\end{eqnarray}
where
$S_{\mathbf{ij}}^{\eta}=\sum_{\alpha\beta}c_{\mathbf{i}\alpha}^{\dag}
\hat{\sigma}_{\alpha\beta}^{\eta}c_{\mathbf{j}\beta}$. Here,
$\hat{\sigma}^{0}$ is the $2\times 2$ identity matrix and
$\hat{\sigma}^{1,2,3}$ are the Pauli matrices. In the Hartree-Fock
approximation, we decouple the operator product
$S_{\mathbf{ij}}^{\eta}(S_{\mathbf{ij}}^{\eta})^{\dag}$ with
$\langle
S_{\mathbf{ij}}^{\eta}\rangle(S_{\mathbf{ij}}^{\eta})^{\dag}
+S_{\mathbf{ij}}^{\eta}\langle(S_{\mathbf{ij}}^{\eta})^{\dag}\rangle
-\langle
S_{\mathbf{ij}}^{\eta}\rangle\langle(S_{\mathbf{ij}}^{\eta})^{\dag}\rangle$.
The expectation value $\langle S_{\mathbf{ij}}^{\eta}\rangle$
defines a four-component vector
\begin{eqnarray}
\langle
S_{\mathbf{ij}}^{\eta}\rangle=&&(\langle\chi_{\mathbf{ij}}\rangle,
\langle\mathbf{S}_{\mathbf{ij}}\rangle),
\end{eqnarray}
where the mean-field amplitudes $\langle\chi_{\mathbf{ij}}\rangle$
and
$\langle\mathbf{S}_{\mathbf{ij}}\rangle=\langle\sum_{\eta=1}^{3}S_{\mathbf{ij}}^{\eta}\rangle$
correspond respectively to the currents in the charge and spin
channels. For the vanadium-based kagome superconductors, only the
charge order is relevant. Thus we need to deal with the case that
$\langle\mathbf{S}_{\mathbf{ij}}\rangle=0$, and this leads to the
mean-field decoupling of the NN Coulomb interaction in the charge
channel as
\begin{eqnarray}
H_{V,MF}=&&-\frac{V_{c}}{2}\sum_{\mathbf{ij}}(\langle\chi_{\mathbf{ij}}^{\dag}\rangle\chi_{\mathbf{ij}}
+\langle\chi_{\mathbf{ij}}\rangle\chi_{\mathbf{ij}}^{\dag}-|\langle\chi_{\mathbf{ij}}\rangle|^{2}
)\nonumber\\
&&+2V_{c}\sum_{\mathbf{i}}n_{\mathbf{i}}.
\end{eqnarray}

In this work, we focus on the CDW states with time reversal symmetry
breaking described by the imaginary part of the mean-field value of
$\chi_{\mathbf{ij}}$ ($\chi_{\mathbf{ij}}^{\dag}$). Using the fact
that
$\sum_{\mathbf{ij}}\langle\chi_{\mathbf{ij}}^{\dag}\rangle\chi_{\mathbf{ij}}
=\sum_{\mathbf{ij}}\langle\chi_{\mathbf{ij}}\rangle\chi_{\mathbf{ij}}^{\dag}$,
we finally arrive at the effective Hamiltonian in Eq. (4). In the
procedure for obtaining Eq. (4), we also neglect the constant term
$\sum_{\mathbf{ij}}|\langle\chi_{\mathbf{ij}}\rangle|^{2}$ and
absorb the term $2V_{c}\sum_{\mathbf{i}}n_{\mathbf{i}}$ into the
chemical potential.

The third term accounts for the SC pairing. It reads
\begin{eqnarray}
H_{P}&=&\sum_{\mathbf{i}}(\Delta
c^{\dag}_{\mathbf{i}\uparrow}c^{\dag}_{\mathbf{i}\downarrow} +h.c.).
\end{eqnarray}
Here, we choose the on-site $s$-wave SC order parameter
$\Delta=-V_{s}\langle
c_{\textbf{i}\downarrow}c_{\textbf{i}\uparrow}\rangle=V_{s}\langle
c_{\textbf{i}\uparrow}c_{\textbf{i}\downarrow}\rangle$. In the
calculations, we choose the typical value of the effective pairing
interaction $V_{s}=1.4$. Varying the pairing interaction will alter
the pairing amplitude, but the results presented here will be
qualitatively unchanged if the strength of CDW order changes
accordingly.

In the coexistence of SC and orbital current orders, the total
Hamiltonian $H=H_{0}+H_{P}+H_{C}$ can be written in the momentum
space within one EUC as,
\begin{eqnarray}
H(\mathbf{k})=&&-t\sum_{\mathbf{k},\langle\mathbf{\tilde{i}\tilde{j}}\rangle,\sigma}c^{\dag}_{\mathbf{k}\mathbf{\tilde{i}}\sigma}c_{\mathbf{k}\mathbf{\tilde{j}}\sigma}e^{-i\mathbf{k}\cdot(\mathbf{r}_{\tilde{i}}-\mathbf{r}_{\tilde{j}})}
-\mu\sum_{\mathbf{k},\mathbf{\tilde{i}},\sigma}c^{\dag}_{\mathbf{k}\mathbf{\tilde{i}}\sigma}c_{\mathbf{k}\mathbf{\tilde{i}}\sigma} \nonumber\\
\nonumber\\
&&+\sum_{\mathbf{k},\langle\mathbf{\tilde{i}\tilde{j}}\rangle,\sigma}iW_{\mathbf{\tilde{i}\tilde{j}}}c^{\dag}_{\mathbf{k}\mathbf{\tilde{i}}\sigma}c_{\mathbf{k}\mathbf{\tilde{j}}\sigma}e^{-i\mathbf{k}\cdot(\mathbf{r}_{\tilde{i}}-\mathbf{r}_{\tilde{j}})}
\nonumber\\
&&+\sum_{\mathbf{k},\mathbf{\tilde{i}}}(\Delta
c^{\dag}_{\mathbf{k}\mathbf{\tilde{i}}\uparrow}c^{\dag}_{-\mathbf{k}\mathbf{\tilde{i}}\downarrow}
+h.c.),
\end{eqnarray}
where $\mathbf{\tilde{i}}\in \textmd{EUC}$ represents the lattice
site being within one EUC, and
$\langle\mathbf{\tilde{i}\tilde{j}}\rangle$ denotes the NN bonds
with the periodic boundary condition implicitly assumed.

\vspace*{.2cm}

Based on the Bogoliubov transformation, we obtain the following
Bogoliubov-de Gennes equations in the EUC,
\begin{eqnarray}
\sum_{\mathbf{k}\mathbf{\tilde{j}}}\left(
\begin{array}{lr}
H_{\mathbf{\tilde{i}\tilde{j}},\sigma} &
\Delta_{\mathbf{\tilde{i}\tilde{j}}} \\
\Delta^{\ast}_{\mathbf{\tilde{i}\tilde{j}}} &
-H^{\ast}_{\mathbf{\tilde{i}\tilde{j}},\bar{\sigma}}
\end{array}
\right)\exp[i\mathbf{k}\cdot(\textbf{r}_{\mathbf{\tilde{j}}}-\textbf{r}_{\mathbf{\tilde{i}}})]\left(
\begin{array}{lr}
u^{\mathbf{k}}_{n,\mathbf{\tilde{j}},\sigma} \\
v^{\mathbf{k}}_{n,\mathbf{\tilde{j}},\bar{\sigma}}
\end{array}
\right) \nonumber\\
= E^{\mathbf{k}}_{n}\left(
\begin{array}{lr}
u^{\mathbf{k}}_{n,\mathbf{\tilde{i}},\sigma} \\
v^{\mathbf{k}}_{n,\mathbf{\tilde{i}},\bar{\sigma}}
\end{array}
\right),
\end{eqnarray}
where
$H_{\mathbf{\tilde{i}}\mathbf{\tilde{j}},\sigma}=(-t+iW_{\mathbf{\tilde{i}\tilde{j}}})\delta_{\mathbf{\tilde{i}}+\mathbf{\tau}_{\mathbf{\tilde{j}}},\mathbf{\tilde{j}}}-\mu\delta_{\mathbf{\tilde{i}},\mathbf{\tilde{j}}}$
with $\mathbf{\tau}_{\mathbf{\tilde{j}}}$ denoting the four NN
vectors and
$\Delta_{\mathbf{\tilde{i}\tilde{j}}}=\Delta\delta_{\mathbf{\tilde{i}},\mathbf{\tilde{j}}}$.
$u^{\mathbf{k}}_{n,\mathbf{\tilde{i}},\sigma}$ and
$v^{\mathbf{k}}_{n,\mathbf{\tilde{i}},\bar{\sigma}}$ are the
Bogoliubov quasiparticle amplitudes on the $\mathbf{\tilde{i}}$-th
site with momentum $\mathbf{k}$ and eigenvalue $E^{\mathbf{k}}_{n}$.
The amplitudes of the SC pairing and the orbital current order, as
well as the electron densities, are obtained through the following
self-consistent equations,
\begin{eqnarray}
\Delta=&&\frac{V_{s}}{2}\sum_{\mathbf{k},n}u^{\mathbf{k}}_{n,\mathbf{\tilde{i}},\sigma}
v^{\mathbf{k}\ast}_{n,\mathbf{\tilde{i}},\bar{\sigma}}
\tanh(\frac{E^{\mathbf{k}}_{n}}{2k_{B}T}) \nonumber\\
W_{\mathbf{\tilde{i}}\mathbf{\tilde{j}}}=&&\frac{V_{c}}{2}\textmd{Im}\{\sum_{\mathbf{k},n}
(u^{\mathbf{k}}_{n,\mathbf{\tilde{i}},\sigma}
u^{\mathbf{k}\ast}_{n,\mathbf{\tilde{j}},\sigma}+v^{\mathbf{k}}_{n,\mathbf{\tilde{i}},\bar{\sigma}}
v^{\mathbf{k}\ast}_{n,\mathbf{\tilde{j}},\bar{\sigma}})\nonumber\\
&&\times\exp[-i\mathbf{k}\cdot(\textbf{r}_{\mathbf{\tilde{j}}}-\textbf{r}_{\mathbf{\tilde{i}}})]
\tanh(\frac{E^{\mathbf{k}}_{n}}{2k_{B}T})\} \nonumber\\
n_{\mathbf{\tilde{i}}}=&&\sum_{\mathbf{k},n}\{|u^{\mathbf{k}}_{n,\mathbf{\tilde{i}},\uparrow}|^{2}f(E^{\mathbf{k}}_{n})+|v^{\mathbf{k}}_{n,\mathbf{\tilde{i}},\downarrow}|^{2}[1-f(E^{\mathbf{k}}_{n})]\}.
\end{eqnarray}

Due to the fairly good FS nesting, the  proximity to the VHF, and
the presence of multiple electronical orders, the self-consistent
calculations may yield several solutions with local energy minima at
the same temperature and doping. In cases where multiple solutions
arise from the self-consistent calculations at the same temperature
but with different sets of initially random input parameters, we
compare their free energy defined as
\begin{eqnarray}
F=&&-2k_{B}T\sum_{\mathbf{k},n,E^{\mathbf{k}}_{n}>0}\ln[2\cosh(\frac{E^{\mathbf{k}}_{n}}{2k_{B}T})]
+N\frac{|\Delta|^{2}}{V_{s}} \nonumber\\
&&+\sum_{\mathbf{k},\langle\mathbf{\tilde{i}}\mathbf{\tilde{j}}\rangle}
\frac{|W_{\mathbf{\tilde{i}}\mathbf{\tilde{j}}}|^{2}}{2V_{c}},
\end{eqnarray}
so as to find the  most favorable state in energy.

Then, the single-particle Green's functions
$G_{\mathbf{\tilde{i}}\mathbf{\tilde{j}}}(\mathbf{k},i\omega)=-\int^{\beta}_{0}d\tau\exp^{i\omega\tau}\langle
T_{\tau}c_{\mathbf{k}\mathbf{\tilde{i}}}(i\tau)c^{\dag}_{\mathbf{k}\mathbf{\tilde{j}}}(0)\rangle$
can be expressed as
\begin{eqnarray}
G_{\mathbf{\tilde{i}}\mathbf{\tilde{j}}}(\mathbf{k},i\omega)=\sum_{n}
\left(\frac{u^{\mathbf{k}}_{n,\mathbf{\tilde{i}},\uparrow}
u^{\mathbf{k}\ast}_{n,\mathbf{\tilde{j}},\uparrow}}{i\omega-E^{\mathbf{k}}_{n}}
+\frac{v^{\mathbf{k}}_{n,\mathbf{\tilde{i}},\downarrow}
v^{\mathbf{k}\ast}_{n,\mathbf{\tilde{j}},\downarrow}}{i\omega+E^{\mathbf{k}}_{n}}\right).
\end{eqnarray}
The spectral function $A(\mathbf{k},E)$ and the DOS $\rho(E)$ can be
derived from the analytic continuation of the Green's function as,
\begin{eqnarray}
A(\mathbf{k},E)=-\frac{1}{N_{P}\pi}\sum_{\mathbf{\tilde{i}}}\textmd{Im}
G_{\mathbf{\tilde{i}}\mathbf{\tilde{i}}}(\mathbf{k},iE\rightarrow
E+i0^{+}),
\end{eqnarray}
and
\begin{eqnarray}
\rho(E)=\frac{1}{N_{\mathbf{k}}}\sum_{\mathbf{k}}A(\mathbf{k},E),
\end{eqnarray}
where $N_{P}$ and $N_{\mathbf{k}}$ are the number of PUCs in the EUC
and the number of $\mathbf{k}$-points in the Brillouin zone,
respectively.

\section{Results and discussion}
\subsection{Phase Diagram}

In the following analysis, the chemical potential $\mu$ is adjusted
to achieve the desired filling. Right at the VHF, the FS possesses a
hexagonal shape, with the saddle points $M_{A/B/C}$ located exactly
on the FS. This unique FS possesses a perfect nesting property and
facilitates the inter-scatterings between three van Hove
singularities connected by the nesting vectors
$\mathbf{Q}_{a}=(-\pi,\sqrt{3}\pi)$,
$\mathbf{Q}_{b}=(-\pi,-\sqrt{3}\pi)$ and $\mathbf{Q}_{c}=(2\pi,0)$,
which has been considered as the primary factors promoting the
so-called ``triple-$Q$" $2\times 2$ CDW in
AV$_{3}$Sb$_{5}$~\cite{Ortiz1,YXJiang1,HZhao1,Liang1,HChen1,HLi2,Mielke1}.
Away from the VHF, the FS becomes more rounded, and the nesting is
weakened, particularly around the saddle points [$M_{A/B/C}$ in
Fig.~\ref{fig1}(b)]. We focus on the situation where the hole doping
is deceased from $1/6$, such that the saddle points move slightly
below the Fermi level as displayed in Figs.~\ref{fig1}(c) and (d),
being consistent with the density functional theory
calculations~\cite{TPark1,JZhao1,YLi1,HZhao1}.

\vspace*{.2cm}
\begin{figure}
\begin{center}
\vspace{.2cm}
\includegraphics[width=230pt,height=230pt]{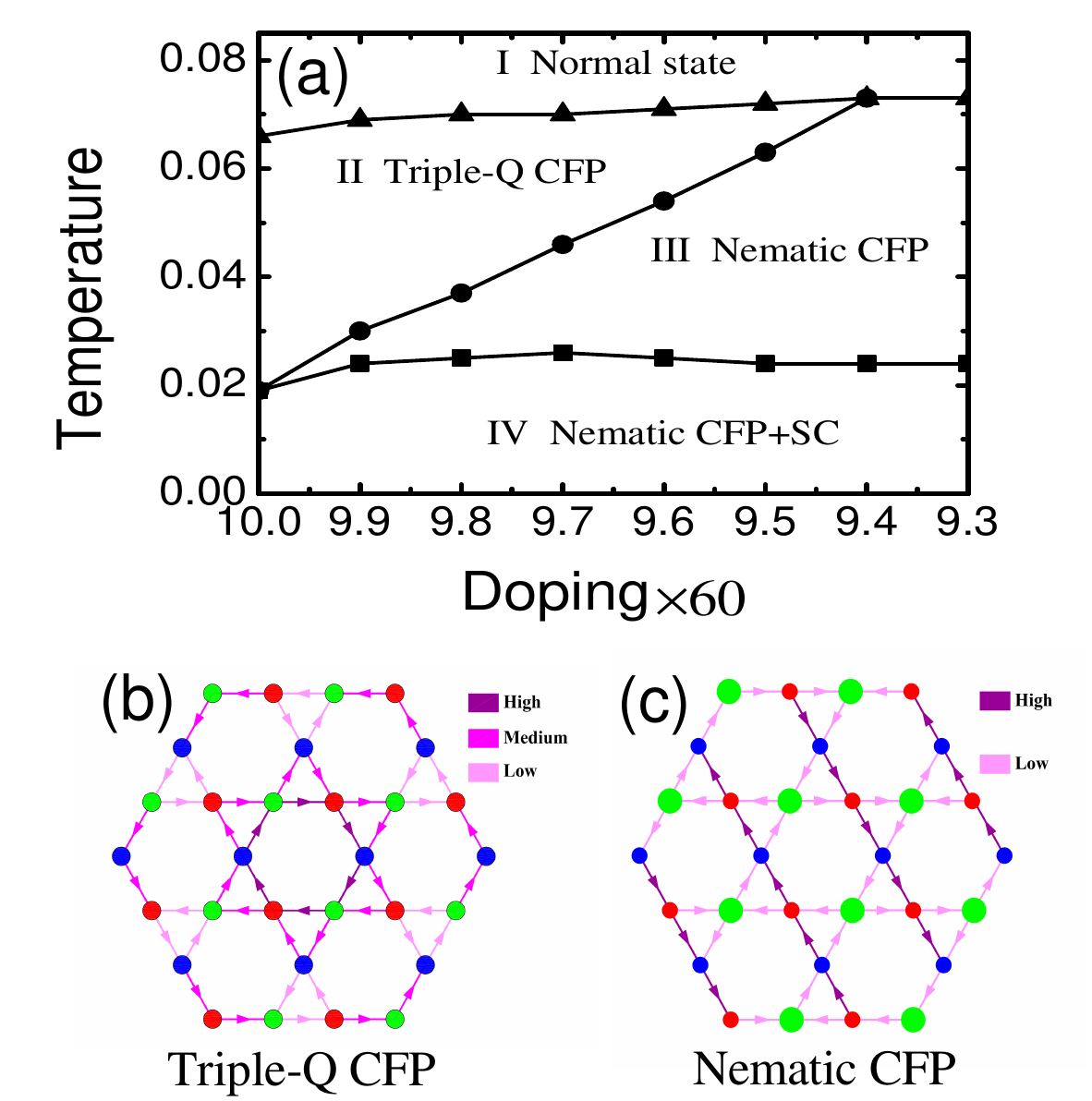}
\caption{(a) Phase diagram as a function of doping and temperature.
The orbital current configurations for the ``triple-$Q$" CFP (b),
and the nematic CFP (c), respectively. The arrows in (b) with their
respective colors indicated by the labels of ``High", ``Medium" and
``Low" signal the magnitudes of the bond current orders. The arrows
in (c) with their respective colors indicated by the labels of
``High" and ``Low" signal the magnitudes of the bond current orders.
In (c), the large (green dots) and small (red and blue dots) sizes
of lattice site also signal respectively the ``High" and ``Low"
values of the on-site SC pairing amplitude in the coexisting
phase(see text and Table II for reference).}\label{fig2}
\end{center}
\end{figure}

As a function of doping and temperature, we find a rich phase
diagram of the model at and near the VHF, which is summarized in
Fig.~\ref{fig2}(a) for the typical values of $V_{s}=1.4$ and
$V_{c}=1.2$. Right at the VHF for $1/6$ doping, where the DOS at the
Fermi level is maximally enhanced and the nesting features of the FS
are strongest, the system prefers the TCFP [Fig.~\ref{fig2}(b)]
before entering the SC state. Once the system deviates from the VHF,
the NCFP order [Fig.~\ref{fig2}(c)] develops in between the TCFP and
the low-temperature SC state. In this case, when decreasing the
temperature, the system starts from the high-temperature normal
state and passes through the NCFP state, and finally transits into
the SC state. When the filling further departs from the van Hove
point, the region of the NCFP expands gradually towards higher
temperatures with the concomitant shrinking of the TCFP region, and
eventually the TCFP is completely displaced by the NCFP state.
Interestingly, the SC state always coexists with the NCFP state in
the low temperature region of the phase diagram, with its free
energy being significantly lower than those of the pure states.
Although the variations of $V_{s}$ and $V_{c}$ may affect the phase
boundaries, the essential feature of the phase diagram, namely the
consecutive evolvement of different ordered states with temperature,
remains qualitatively unchanged.

It is remarkable that the successive temperature evolutions from the
TCFP phase to the NCFP state, occurring at a doping level slightly
deviating from the van Hove point in a self-consistent manner,
exhibits the same trend as the experimental
observations~\cite{HLi1,ZJiang1,LNie1,PWu1}. Particularly, the
ground state characterized by the coexistence of the NCFP and SC
orders may be related to the $C_{2}$ symmetry and the
time-reversal-symmetry breaking observed in the SC
state~\cite{Xiang1,HZhao1,HLi1}.

\begin{figure}
\begin{center}
\vspace{.2cm}
\includegraphics[width=240pt,height=90pt]{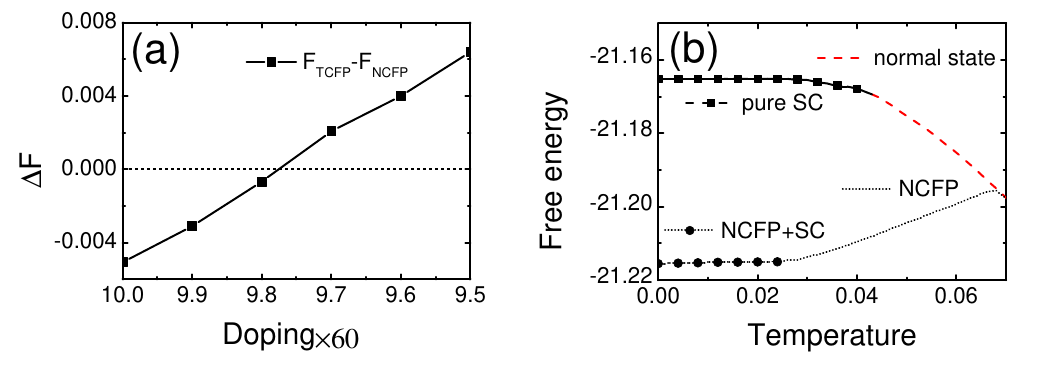}
\caption{(a) Doping dependence of the free energy difference per
site between the ``triple-$Q$" CFP and nematic CFP. In obtaining the
results in (a), the CDW states are each intentionally kept in their
respective forms during the calculations. (b) The evolutions of the
free energy per site from the normal state (dashed line) to pure SC
state (dashed ling with square symbol), and from the nematic CFP
(dotted line) to the coexisting phase with nematic CFP and SC state
(dotted line with round symbol) at doping level
$9.8/60$.}\label{fig3}
\end{center}
\end{figure}

Since the effect of doping on the phase diagram is closely related
to the VHF, we will focus on the physics associated with the three
van Hove points, in addition to the nesting properties of the FS. At
each van Hove point, the electronic states come exclusively from one
of the three distinct sublattices~\cite{SLYu1}. As a result, the
scattering between low-energy electronic states connected by each
nesting wave vector occurs solely between two sublattices. This
unique property creates the necessary conditions for the transition
from the TCFP to the NCFP through doping. At VHF, the electronic
states at the three van Hove points are mutually coupled by the CDW
orders with three wave vectors $\mathbf{Q}_{a}$, $\mathbf{Q}_{b}$
and $\mathbf{Q}_{c}$ in the TCFP in an end-to-end manner
[Fig.~\ref{fig1}(b)]. This is to say, the CDW orders with three wave
vectors are mutually coupled in pairs with the strongest coupling
strength at the van Hove points. Thus, as depicted in the phase
diagram [Fig.~\ref{fig2}(a)], a stable charge order pattern that
simultaneously satisfies the three wave vectors can be found at the
VHF. However, when the system deviates from the VHF by reducing the
hole doping, the chemical potential $\mu$ is elevated, and
accordingly the saddle points move below the Fermi level, as
demonstrated in Figs.~\ref{fig1}(c) and (d). The deviation of the FS
from the saddle points weakens the mutual couplings between pairs of
the three wave vectors and, correspondingly, the TCFP. In this
situation, there is a significant decrease in the energy difference
between the TCFP, which satisfies three ordered wave vectors, and
the NCFP, which has only one ordered wave vector. Moreover, the NCFP
will deform the FS, suppressing the other two NCFP with different
ordered wave vectors while further enhancing itself [refer to
Fig.~\ref{fig2}(c) and Table I]. Consequently, within a certain
doping range, the NCFP becomes more stable than the TCFP. For better
clarity, we present the evolutions of the free energy deference
between the TCFP and the NCFP with doping at a specific temperature
$T=0.04$ in Fig.~\ref{fig3}(a). It shows that the TCFP has a lower
free energy than that of the NCFP, when the doping level has not
much deviations from the VHF. Nevertheless, as the system deviates
appreciably from the VHF, the NCFP acquires the lower free energy.
As a result, a spontaneous rotational symmetry-breaking transition
occurs from the TCFP to the NCFP at the doping level defined by the
zero point of the free energy difference.

\vspace*{.2cm}

The temperature effects on the CDW states are also closely related
to the van Hove physics. Although the van Hove points shift below
the Fermi level for doping levels deviating from the VHF, the
thermal broadening effect becomes prominent at relatively high
temperatures, thereby increasing the effectiveness of the van Hove
points. This enhances the mutual coupling among three CDW orders
associated with different wave vectors $\mathbf{Q}_{a}$,
$\mathbf{Q}_{b}$ and $\mathbf{Q}_{c}$. As a result, for doping
levels that deviate from the VHF, the TCFP is stabilized by the
thermal broadening effect at relatively high temperatures, while the
NCFP becomes more favorable due to the reduction of thermal
broadening effect at low temperatures.

Previously, the transition to the CDW with $C_{2}$ symmetry was
proposed to arise from interlayer interactions between adjacent
kagome planes with already existing $C_{6}$-symmetry charge orders
in each single layer~\cite{TPark1,Christ1}. As a secondary outcome
of the $C_{6}$-symmetry charge orders in this interlayer coupling
scenario, the nematicity typically occurs at a lower temperature
$T_{nem}$ well below $T_{CDW}$. However, as shown in
Fig.~\ref{fig2}(a), our theory shows a regime of less than $9.4/60$
hole doping where the NCFP directly straddles the normal and SC
states, despite in the high doping level regime of the phase diagram
the appearance of nematic CDW at low temperatures is well below
$T_{CDW}$. Interestingly, a recent experiment has indeed observed an
immediate development of nematicity and possible time-reversal
symmetry breaking in the CDW state of CsV$_{3}$Sb$_{5}$~\cite{QWu1},
providing further support on our theory.

As the temperature continues to decrease, several ingredients
promote the development of the coexisting phase of the NCFP and SC
state. First of all, as depicted in Figs.~\ref{fig3}(b) and
~\ref{fig6}(a), the NCFP exhibits a lower energy compared to the
normal and TCFP states before the SC transition, making it
energetically favorable as the parent state for the formation of the
SC order. Secondly, the well-preserved portions of the FS,
especially those portions near the saddle points [the $M_{A}$ points
in Fig~\ref{fig4}(b)], within the NCFP provide sufficient electronic
states for the formation of the SC pairing. Thirdly, apart from the
gaped portions of the FS, the one-wave-vector scattering with the
rotational symmetry breaking in the NCFP state induces a slight
deformation of the  FS [refer to Figs.~\ref{fig4}(b) and
~\ref{fig6}(b)], which brings the remaining FS closer to the van
Hove points compared to the normal state. As a result, the
coexisting phase of NCFP and SC orders possesses the significantly
lower free energy than that for the pure SC state [see
Fig.~\ref{fig3}(b)].

\subsection{Characteristics of electronic states}

Next, we will investigate in detail on the electronic structures in
different regime of the phase diagram, namely the TCFP state, the
NCFP state, and the coexisting state of the NCFP and SC orders. To
demonstrate our results, we will focus on the typical cases with a
doping level of $9.8/60$ at temperatures $T=0.05$, $T=0.03$ and
$T=1\times 10^{-5}$, corresponding to the TCFP state, the NCFP state
and the coexisting phase of the NCFP and SC orders, respectively.

\begin{figure}
\begin{center}
\vspace{.2cm}
\includegraphics[width=240pt,height=200pt]{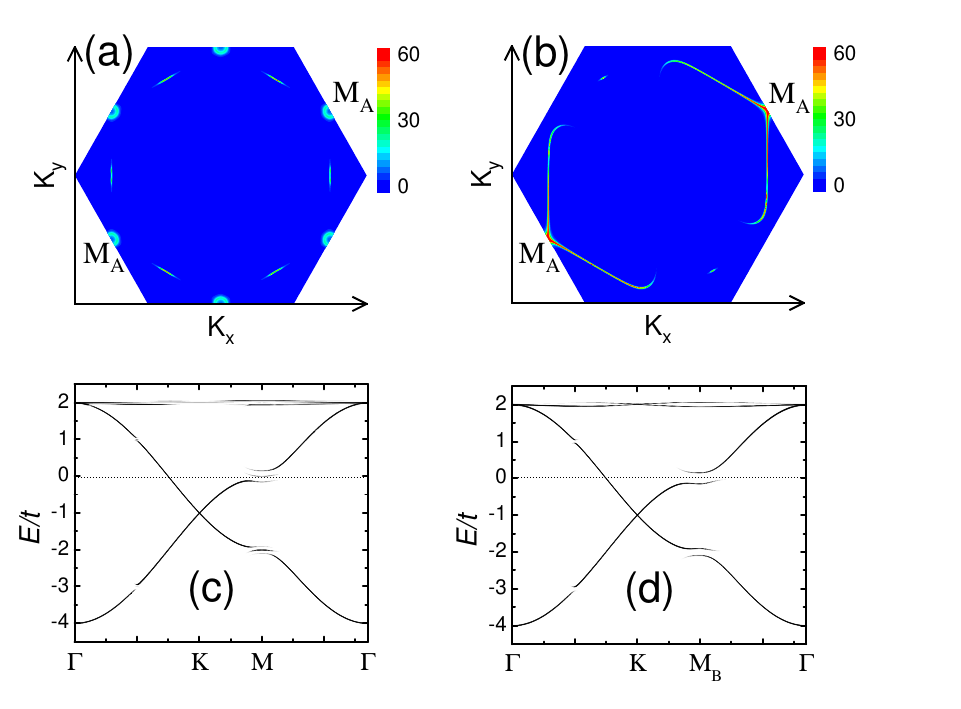}
\caption{Zero-energy spectral weight distribution $A(\mathbf{k},E)$
unfolded in the primitive Brillouin zone for hole doping $9.8/60$ at
$T=0.05$ (a), and at $T=0.03$ (b), respectively. The former situates
in the ``triple-$Q$" CFP region, while the latter lies in the region
of the nematic CFP state. The unfolded dispersions along
high-symmetry cuts in the primitive Brillouin zone are shown
correspondingly in (c) for the ``triple-$Q$" CFP state, and in (d)
for the nematic CFP state, respectively.}\label{fig4}
\end{center}
\end{figure}

\begin{figure}
\begin{center}
\vspace{.2cm}
\includegraphics[width=240pt,height=190pt]{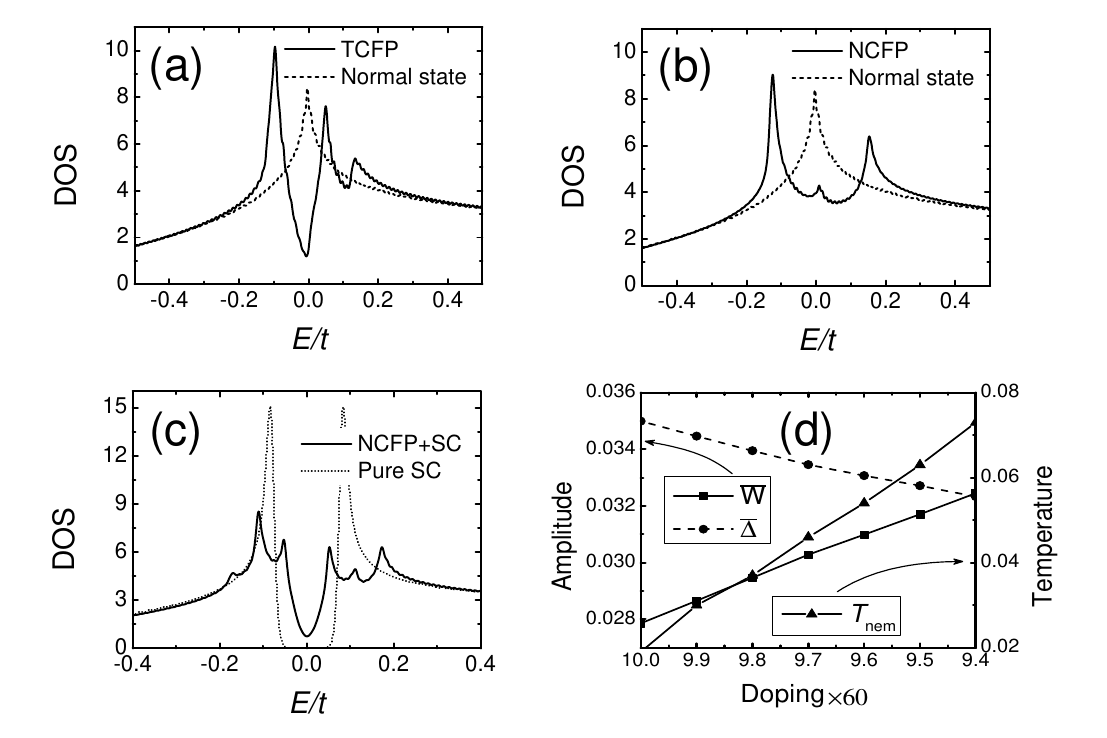}
\caption{Energy dependence of the DOSs for the ``triple-$Q$" CFP
(a), the nematic CFP (b), and the coexisting phase of the nematic
CFP and SC orders (c), respectively. (d) The doping dependence of
the averaged SC pairing amplitude $\bar{\Delta}$, the averaged
magnitude of the bond current orders $\bar{W}$, and the transition
temperature of the nematic CFP $T_{nem}$.}\label{fig5}
\end{center}
\end{figure}

For the TCFP, the bond current orders obtained from self-consistent
calculations can be classified into three levels of magnitude:
``High", ``Medium" and ``Low", as displayed in Fig.~\ref{fig2}(b)
and listed in Table I. These orders give rise to a special pattern
known as the ``Star of David" state. Fig.~\ref{fig4}(a) presents the
distribution of spectral weight $A(\mathbf{k},E)$ at $E=0$, which is
unfolded in the primitive Brillouin zone. Consistent with the
previous non-self-consistent results~\cite{HMJiang2}, the
zero-energy spectral weight distribution clearly reveals partially
gapped Fermi segments and preserves the $C_{6}$ symmetry.
Correspondingly, the DOS shown in Fig.~\ref{fig5}(a) exhibits a
``pseudogap-like" feature, with non-zero minima occurring at $E=0$.
In contrast, the DOS for the normal state, represented by the black
dashed curve in the same figure, displays the typical van Hove peak
near $E=0$.

\vspace*{.2cm}

On the other hand, as the temperature decreases into the NCFP
regime, such as $T=0.03$, the magnitude distribution of the three
inequivalent bond current orders transforms into that of two
inequivalent orders. The ``Low" intensity is associated with two
inequivalent bonds, which are approximately one order of magnitude
smaller than the other one labeled as ``High" intensity. In this
case, only the strong current order between sublattices $B$ and $C$
corresponds to the charge order scattering between $M_{B}$ and
$M_{C}$ points with a momentum transfer $\mathbf{Q}_{a}$. As a
result, the depletion of the zero-energy spectral weight only occurs
in regions connected by one of the three wave vectors, such as
$\mathbf{Q}_{a}$ in Fig.~\ref{fig4}(b), manifesting the
characteristics of the $C_{2}$ symmetry. The DOS for the NCFP
depicted by the black solid curve in Fig.~\ref{fig5}(b) still
exhibits two peaks resembling gap edges, but a significant DOS shows
up within the peak edges, accompanied by a residue van Hove peak
near the zero bias, resulting from the large portion of the unspoilt
Fermi segments and the preserved van Hove points $M_{A}$.

\begin{table}[]
\begin{center}
\begin{tabular}{|c|c|c|c|}
\hline
                                 & High & Medium & Low \\ \hline
 {``triple-$Q$" CFP} & 0.043805 & 0.035128 & 0.026666 \\ \hline 

{nematic CFP} & 0.071768 &  & 0.00841  \\ \hline
\end{tabular}
\end{center}
\caption{Magnitude of the bond current order $|W_{ij}|$ in the
self-consistent calculations at filling level $9.8/60$ with $T=0.05$
for the ``triple-$Q$" CFP and $T=0.03$ for the nematic CFP.}
\end{table}

Let's further analyze the characteristics of the band structures in
the two CDW states. In the TCFP, the bands along different
high-symmetry cuts $\Gamma\rightarrow K/K'\rightarrow
M_{A/B/C}\rightarrow\Gamma$ remain the same due to the preservation
of the $C_{6}$ symmetry, while in the NCFP, a gap opens near the
$M_{B/C}$ points but not near the $M_{A}$ point, as shown in
Figs.~\ref{fig4}(d), ~\ref{fig7}(a) and ~\ref{fig7}(b). In addition,
near the saddle point, the band is triply split in the TCFP, while
it is only double split near the $M_{B/C}$ points in the NCFP, as
illustrated in Figs.~\ref{fig4}(c), (d) and Fig. ~\ref{fig7}(a).
This behavior can be understood through the ``patch model", which
provides an approximate description of the low-energy scatterings
between the saddle points~\cite{Nand1,YPLin1,YPLin2}. In the TCFP
state, the patch model involving the ``triple-$Q$" scatterings
reads,
\begin{eqnarray}
H_{TCFP}(M)=\left(
\begin{array}{ccc}
\varepsilon_{M_{A}} & i\lambda_{AB} &
i\lambda_{AC} \\
-i\lambda_{AB} & \varepsilon_{M_{B}} & i\lambda_{BC} \\
-i\lambda_{AC} & -i\lambda_{BC} & \varepsilon_{M_{C}}
\end{array}
\right).
\end{eqnarray}
Nevertheless, the charge order scattering in the NCFP involves only
one wave vector, such as $\mathbf{Q}_{a}$ that corresponds to the
bond current configuration in Fig.~\ref{fig2}(c). As a result, the
patch model in the NCFP state is reduced to
\begin{eqnarray}
H_{NCFP}(M)=\left(
\begin{array}{cc}
\varepsilon_{M_{B}} & i\lambda_{BC} \\
-i\lambda_{BC} & \varepsilon_{M_{C}}
\end{array}
\right).
\end{eqnarray}
Here, $\varepsilon_{M_{A}}$ ($\varepsilon_{M_{B}}$,
$\varepsilon_{M_{C}}$) stands for the energy at the saddle point
$M_{A}$ ($M_{B}$, $M_{C}$) that originates from the sublattice $A$
($B$, $C$), and $\lambda_{AB}$ ($\lambda_{AC}$, $\lambda_{BC}$)
represents the scattering strength of the bond current order between
$M_{A}$ and $M_{B}$ ($M_{A}$ and $M_{C}$, $M_{B}$ and $M_{C}$). Near
the VHF, where
$\varepsilon_{M_{A}}=\varepsilon_{M_{B}}=\varepsilon_{M_{C}}\approx
0$, one can immediately find that the Hamiltonian $H_{TCFP}(M)$ has
three eigenvalues $E_{0}(M)=0$ and
$E_{\pm}(M)=\pm\sqrt{\lambda_{AB}^{2}+\lambda_{AC}^{2}+\lambda_{BC}^{2}}$.
The Hamiltonian $H_{NCFP}(M)$, on the other hand, has two
eigenvalues $E_{\pm}(M)=\pm|\lambda_{BC}|$. It is worth pointing out
that the unique change of the electronic structures from the TCFP to
the NCFP can serve as an indirect evidence to identify the
electronic nematicity in AV$_{3}$Sb$_{5}$.

\begin{table}[]
\begin{center}
\begin{tabular}{|c|c|c|}
\hline
                                 & High & Low \\ \hline
{nematic CFP} & 0.072113 & 0.008144 \\ \hline
{SC} & 0.050711 & 0.025567  \\ \hline
\end{tabular}
\end{center}
\caption{Magnitudes of the bond current order $|W_{ij}|$ and the SC
pairing $|\Delta|$ in the self-consistent calculations at filling
level $9.8/60$ with $T=1\times 10^{-5}$ for the coexisting phase of
the nematic CFP and SC orders.}
\end{table}

Then, we turn to the coexisting phase of the NCFP and SC orders. On
the one hand, as indicated in Table II, the strength of the bond
current orders changes little upon entering the coexisting phase. On
the other hand, as displayed in Fig.~\ref{fig2}(c), the distribution
of SC pairing amplitudes depends not only on the strength but also
on the direction of the surrounding bond current orders.
Specifically, the ``High" value of the SC pairing amplitude appears
at the sublattice site [the sublattice $A$ in Fig.~\ref{fig2}(c)]
where the surrounding bond current orders are weak and the bonds
connected to the same sublattice sites carry either the same inflow
current directions or the same outflow current directions. On the
contrary, the ``Low" value of the SC pairing amplitude appears on
the sublattice sites where a pair of bonds connected to the same
sublattice sites have respective inflow and outflow current
directions. The uneven distribution of the SC pairing amplitude can
be understood from the low energy spectral distribution shown in
Fig.~\ref{fig4}(b). In the coexisting phase of the NCFP and SC
orders, since the depletion of the low energy spectral weight only
occurs at portions between $M_{B}$ and $M_{C}$, the remained
spectral weights, including the perfect van Hove points $M_{A}$
caused by the deformation of the FS, mainly come from the sublattice
$A$. Consequently, the SC pairing amplitude on sublattice $A$ is
significantly larger than those on the sublattices $B$ and $C$.

\vspace*{.2cm}

Due to the uneven distributions of the SC pairing amplitude and the
scattering of the CDW order, a V-shaped DOS can be observed in
Fig.~\ref{fig5}(c), accompanied by multiple sets of coherent peaks
and residual zero-energy DOS, constituting a characteristic of a
nodal multi-gap SC pairing state as having been observed in the STM
experiments~\cite{HSXu1,Liang1,HChen1}. For comparison, the dotted
curve in the same figure portrays a typical U-shaped full gap
structure for the DOS in the state with pure on-site $s$-wave SC
pairing.

Considering the nature of the mean-field approximation, it should be
noted that the transition temperature between different phases in
the calculated results is just qualitative rather than quantitative.
Nevertheless, the anti-correlation trend between the SC pairing
amplitude and the transition temperature of the nematic phase at
different doping levels can be clearly observed in
Fig.~\ref{fig5}(d), as supported by the STM experiments~\cite{PWu1}.

\section{Conclusion}
In conclusion, we have investigated the origin of the chiral CDW and
its interplay with superconductivity in a fully self-consistent
theory considering the orbital current order and the on-site SC
pairing, which determines both the CDW and the SC orders
self-consistently. It was revealed that the self-consistent theory
captures the salient feature for the successive temperature
evolutions of the ordered electronic states from the
high-temperature $2\times 2$ TCFP to the NCFP, and to the
low-temperature $s$-wave SC state in a coexisting manner with the
NCFP order. The rotational symmetry breaking transition of the CDW
could be understood from a scenario in which the competition between
the deviation from the VHF and the thermal broadening of the FS
determines which state is it in. The intertwining of the $s$-wave SC
pairing with the NCFP order produced a nodal gap feature manifesting
as the V-shaped DOS along with the residual DOS near the Fermi
energy. The self-consistent theory not only produced the successive
temperature evolutions of the electronically ordered states observed
in experiment, but might also offer a heuristic explanation to the
two-fold rotational symmetry of electron state detected in both the
CDW and the SC states. Moreover, the intertwining of the SC pairing
with the NCFP order, which was found to be a ground state in the
self-consistent theory at the low temperature regime, might also be
a promising alternative for mediating the divergent or seemingly
contradictory experimental outcomes regarding the SC properties.
Overall, our study sheds light on the intricate relationship between
the chiral CDW and superconductivity, providing valuable insights
into the underlying mechanisms and experimental observations.

\section{acknowledgement}
\par
This work was supported by National Key Projects for Research and
Development of China (Grant No. 2021YFA1400400), and the National
Natural Science Foundation of China (Grants No. 12074175, No.
12374137 and No. 92165205).


\appendix*


\renewcommand{\thefigure}{\Alph{section}\arabic{figure}}

\hspace{1.6cm}

\section{Temperature evolution of free energy, details of spectrum and band structures along high-symmetry cuts}

In Fig.~\ref{fig6}(a), we show the temperature evolution of the free
energy in the TCFP and in the NCFP. In Fig.~\ref{fig6}(b), we
display the momentum cut of the spectral weight along the
$M_{A}\rightarrow\Gamma\rightarrow M_{A}$ direction [see Fig. 4(b)
in the main text] at a doping level deviating from the VHF for the
normal state and for the NCFP.

\setcounter{figure}{0}
\begin{figure}[h]
\begin{center}
\vspace{.6cm}
{\mbox{\includegraphics[width=230pt,height=100pt]{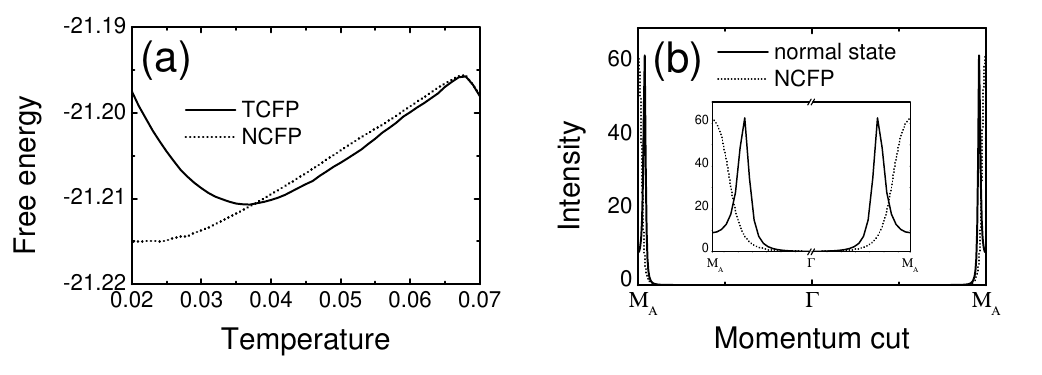}}}
\caption{(a) Temperature evolution of the free energy per site for
the doping level $9.8/60$ in the ``triple-$Q$" CFP (solid line) and
in the nematic CFP (dotted line). In obtaining the results in (a),
the CDW states are intentionally kept in their respective forms
during the calculations. (b) The momentum cut of the spectral weight
along the $M_{A}\rightarrow\Gamma\rightarrow M_{A}$ direction at a
doping level deviating from the VHF for the normal state (solid
curve) and for the nematic CFP (dotted curve). The peaks of the
spectral weight intensity denote the position of the Fermi surface
[refer to Fig. 4(b) in the main text]. The same figure of (b) is
replotted in the inset by breaking the $x$-axis in order to have a
better view of the Fermi surface shift.} \label{fig6}
\end{center}
\end{figure}

\begin{figure}[h]
\begin{center}
{\mbox{\includegraphics[width=230pt,height=100pt]{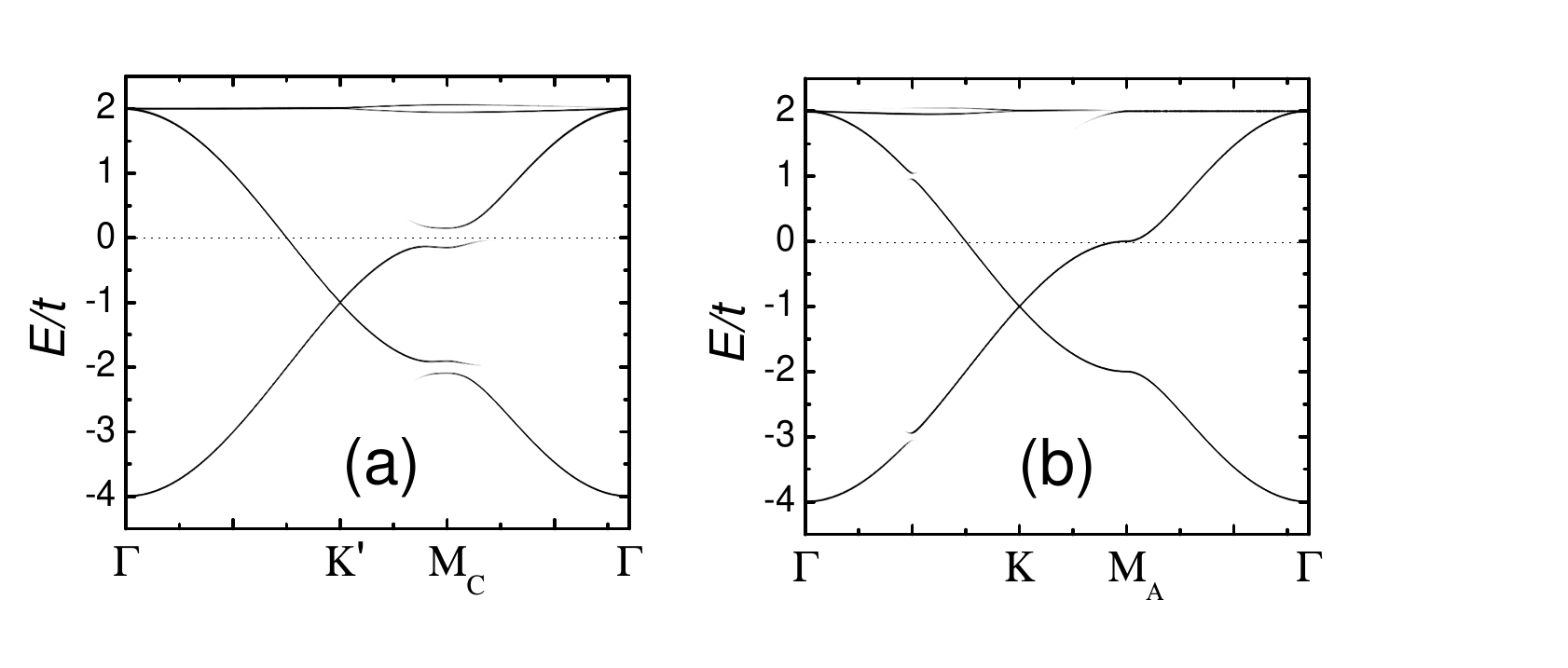}}}
\caption{The unfolded dispersions along high-symmetry cuts
$\Gamma$-$K'$-$M_{C}$-$\Gamma$ (a), and
$\Gamma$-$K$-$M_{A}$-$\Gamma$ (b), respectively.} \label{fig7}
\end{center}
\end{figure}

In Fig.~\ref{fig7}, we present the unfolded dispersions of the
spectral weight along different high-symmetry cuts for the NCFP.
Owing to the $C_{2}$ symmetry of the NCFP, the energy bands exhibit
different features along different high symmetry cut. Specifically,
an energy gap opens near the $M_{B/C}$ points but not near the
$M_{A}$ point, as presented respectively in Figs.~\ref{fig7}(a) and
(b).

\end{document}